\documentclass[preprint2]{aastex}

\shorttitle{The spin-down timescale of SN Ia progenitors} \shortauthors{Meng
\& Podsiadlowski}

\begin{document}


\title{Constraining the spin-down timescale of the white-dwarf progenitors
of Type Ia supernovae}


\author{Xiangcun. Meng}
\affil{School of Physics and Chemistry, Henan Polytechnic
University, Jiaozuo, 454000, China\\
Key Laboratory for the Structure and Evolution of Celestial
Objects, Chinese Academy of Sciences, Kunming 650011, China}
\email{xiangcunmeng@hotmail.com}
\and
\author{Philipp. Podsiadlowski}
\affil{Department of Astronomy, Oxford University, Oxford OX1 3RH} 






\begin{abstract}
Justham (2011) and di\,Stefano et al.\ (2011) proposed that the
white-dwarf progenitor of a Type Ia supernova (SN Ia) may have to
spin down before it can explode. As the white dwarf spin-down
timescale is not well known theoretically, we here try to
constrain it empirically (within the framework of this spin-down
model) for progenitor systems that contain a giant donor and for
which circumbinary material has been detected after the explosion:
we obtain an upper limit of a few $10^{\rm 7} {\rm yr}$. Based on
the study of \citet{DISTEFANO12}, this means that it is too early
to rule out the existence of a surviving companion in SNR
0509-67.5.

\end{abstract}


\keywords{binaries: symbiotic - stars: evolution- star: mass-loss
- supernovae: general - star: white dwarfs}



\section{INTRODUCTION}\label{sect:1}
Type Ia supernovae (SNe Ia) appear to be good cosmological distance
indicators and have been successfully used for determining
cosmological parameters (e.g. $\Omega$ and $\Lambda$); this has led to
the discovery that the Universe is accelerating (\citealt{RIE98};
\citealt{PER99}). It has also been proposed that they can be used to
test the evolution of the dark energy equation of state with time
(\citealt{HOWEL09}). However, the progenitor systems of SNe Ia have
not yet been confidently identified (\citealt{HN00}; \citealt{LEI00}),
although the identification of the progenitor would be important in many
astrophysical fields (\citealt{WANGB12}). Two basic scenarios have
been competing for about four decades. In the single degenerate (SD)
model, a carbon-oxygen white dwarf (CO WD) grows in mass via accretion
from its non-degenerate companion (\citealt{WI73}; \citealt{NTY84}),
while in the double degenerate (DD) scenario, two WDs merge after
losing angular momentum by gravitational wave radiation
(\citealt{IT84}; \citealt{WEB84}).

Searching for the surviving companion in a supernova remnant
(SNR), only predicted in the SD scenario, provides one of the most
promising methods for distinguishing between the two basic
scenarios. The claim of the discovery of a potential surviving SN
companion in the Tycho supernova remnant has been hailed as a
major advance in this field (Tycho G, \citealt{RUI04}), although
serious doubts have been raised that Tycho G actually is the
surviving companion (\citealt{KERZENDORF09, KERZENDORF13};
\citealt{HERNANDEZ09}; \citealt{SHAPPEE13}). Recently,
\citet{SCHAEFER12} searched for a potential surviving companion in
SNR 0509-67.5 and reported a negative result, apparently favoring
the DD model. On the other hand, \citet{DISTEFANO12} argued that,
if the WD experienced a long spin-down phase before the explosion,
the donor star could be too dim to be detectable at the time of
explosion. Unfortunately, the spin-down timescale is very
uncertain (\citealt{DISTEFANO11}), and it is therefore difficult
to arrive at a firm conclusion with regard to the observations by
\citet{SCHAEFER12}.

There has been evidence for circumstellar material (CSM) in the
spectrum of a number of SNe Ia, which is usually taken as evidence
in favor of the SD model (\citealt{PAT07}; \citealt{STERNBERG11};
\citealt{DILDAY12}; \citealt{MAGUIRE13}). The fact that the CSM
can be detected indicates that the companions are still losing
material before the SN Ia explosion or only ceased transferring
mass relatively recently before the explosion. This implies that
the spin-down timescale should be shorter than the timescale on
which the companion loses all of its remaining envelope; otherwise
the signature of the CSM should not be detected. In this paper, we
want to use this idea to empirically constrain the spin-down
timescale (within the framework of the spin-down model).

In Section \ref{sect:2}, we describe our method and present the
results of our calculations in Section \ref{sect:3}. In Section
\ref{sect:4}, we  discuss their implications.

\section{METHOD}
\label{sect:2}

If the WD companion is a main-sequence (MS) star, the wind
mass-loss rate will be relativley low and the velocity of outflow
from the binary system very high; and it is then very difficult to
detect the CSM. Therefore, we only consider WD + red giant (RG)
systems here. Although the details of detecting the CSM from WD +
RG systems depend on the system parameters, the CSM has be
detected in at least some systems (\citealt{PAT07};
\citealt{DILDAY12}). We will address this problem again in another
future paper. The method to follow the binary evolution here is
very similar to the prescriptions presented in \citet{CHENXF11}

We use the stellar evolution code of \citet{EGG71} to calculate
the binary evolutions of SD systems. Roche lobe overflow (RLOF) is
treated within the code as described by \citet{HAN00}. A `standard' solar
metallicity is adopted ($Z=0.02$). The opacity tables for the
metallicity come from the compilations from \citet{CHE07}, \citet{IR96} and
\cite{AF94}.

We also assume that the stellar wind mass-loss rate of the secondary in a
binary system is increased by the presence of the WD companion
star. Specifically, the tidal enhancement of mass-loss rate from secondary
is modelled using the Reimers' (1975) wind-mass formula with an extra tidal
term following \citet{TOUT88}:


 \begin{equation}
 \begin{array}{lc}
 \dot{M}_{\rm 2w}=-4\times10^{\rm -13}{\displaystyle{\eta(L/L_{\odot})(R/R_{\odot})}\over\displaystyle{(M_{\rm
 2}/M_{\odot})}}\\
 \\
 \hspace{0.75cm}\times\left[1+B_{\rm W}\min\left(\frac{1}{2},\frac{R}{R_{\rm
L}}\right)^{\rm 6}\right] M_{\odot}{\rm yr}^{\rm -1},
\end{array}
  \end{equation}
where $L$ and $R$ are the luminosity and the radius of the giant
secondary, $R_{\rm L}$ is its Roche lobe radius, and $\eta=0.25$
is the Reimers' wind coefficient. The wind enhance factor $B_{\rm W}$
is still uncertain; it is more than 3000 in \citet{TOUT88} and
$10^4$ in the wind-driven mass transfer theory of
\citet{TOUT91}.  Here, we set $B_{\rm W}=10000$, which means that
the mass-loss rate from the secondary could be 150 times as large
as the Reimers' rate when the star fills more than half its Roche
lobe.

Some of the material lost in the stellar wind from the
secondary may be accreted by the WD; the resulting mass-accretion rate is
expressed as (from \citealt{BOFFIN88})

 \begin{equation}
 \dot{M}_{\rm 2a}=-\frac{1}{\sqrt{1-e^{\rm 2}}}\left(\frac{GM_{\rm WD}}{v_{\rm w}^{\rm 2}}\right)^{\rm 2}\frac{\alpha_{\rm acc}\dot{M}_{\rm 2w}}{2a^{2}(1+v_{\rm orb}^{\rm 2}/v_{\rm w}^{\rm
 2})^{3/2}},
  \end{equation}
where $v_{\rm orb}=\sqrt{G(M_{\rm 2}+M_{\rm WD})/a}$ is the orbital
velocity, $G$ Newton's gravitational constant, $a$ the semi-major axis
of the orbit, and $e$ its eccentricity. In this paper, we take
$e=0$. The accretion efficiency ($\alpha_{\rm acc}$) is set as
1.5. For simplicity, we set $v_{\rm w}=500 {\rm km}$ ${\rm s}^{\rm
  -1}$ for a MS star and $5 {\rm km}\,{\rm s}^{\rm -1}$ for a
RG. Here, $5 {\rm km}$\,${\rm s}^{\rm -1}$ is a lower limit for the
wind velocity (see also \citealt{CHENXF11}). In this equation, for
sufficiently small $a$, the right-hand side can be larger than
$-\dot{M}_{\rm 2w}$; we therefore limit $\dot{M}_{\rm
  2a}\leq-\dot{M}_{\rm 2w}$, as did \citet{CHENXF11}. In fact, both
$B_{\rm W}$ and $v_{\rm w}$ are poorly known and only upper or
lower limits are used here for $B_{\rm W}$ and $v_{\rm w}$,
respectively. \citet{CHENXF11} have shown that the parameter space
leading to SNe Ia in the orbital period--secondary mass plane
increases with $B_{\rm W}$ and decreases with $v_{\rm
  w}$. Since $B_{\rm W}=10000$ and $v_{\rm w}=5 {\rm km}\,{\rm
    s}^{\rm -1}$ may be considered to be upper and lower limits for
  the two parameters, respectively, the resulting parameter space
leading to SNe Ia in the orbital period--secondary mass plane is
the most generous case and should cover those from other parameter
combinations.  We will discuss their effects on the results, as
well as possible metallicity effects, in section \ref{sect:4.1}.

Wind accretion is the only way to transfer material from the RG
secondary to the WD before Roche lobe overflow (RLOF) begins; then the
mass-transfer rate is $\dot{M}_{\rm tr}=\dot{M}_{\rm 2a}$. After RLOF
has started, material is transferred by both an accretion stream from
the inner Lagrangian point and the wind so that $\dot{M}_{\rm
  tr}=\dot{M}_{\rm 2a}+|\dot{M}_{\rm 2RLOF}|$, where $\dot{M}_{\rm
  2RLOF}$ is the mass-transfer rate by RLOF.

We adopt the prescription of \citet{HAC99a} on WDs accreting
hydrogen-rich material from their companions (see details in
\citealt{HAN04} and \citealt{MENG09}). Then, the mass growth rate
of the CO WD, $\dot{M}_{\rm WD}$, is
 \begin{equation}
 \dot{M}_{\rm WD}=\eta_{\rm He}\eta_{\rm
 H}\dot{M}_{\rm tr},
  \end{equation}
where $\eta _{\rm H}$ is the mass-accumulation efficiency for
hydrogen burning  and is controlled by
 \begin{equation}
\eta _{\rm H}=\left\{
 \begin{array}{ll}
 \dot{M}_{\rm c}/\dot{M}_{\rm tr}, & \dot{M}_{\rm tr}> \dot{M}_{\rm
 c},\\
 1, & \dot{M}_{\rm c}\geq \dot{M}_{\rm tr}\geq\frac{1}{8}\dot{M}_{\rm
 c},\\
 0, & \dot{M}_{\rm tr}< \frac{1}{8}\dot{M}_{\rm c},
\end{array}\right.
\end{equation}
where $\dot{M}_{\rm c}$ is the critical accretion rate for stable
hydrogen burning; $\eta_{\rm He}$ is the mass-accumulation efficiency
for helium-shell flashes, and its value is taken from
\citet{KH04}. The material lost in the form of an optically thick wind
(\citealt{HAC96}) is assumed to take away the specific orbital angular
momentum of the accreting WD, while the wind material from the
secondary that is not accreted by the WD is assumed to take away the
specific orbital angular momentum of the donor star. For simplicity, we
choose $M_{\rm WD}^{\rm i}=1.0\,M_{\odot}$ and $1.1\,M_{\odot}$, while
$M_{\rm 2}^{\rm i}$ ranges from 0.8\,$M_{\odot}$ to 5.6\,$M_{\odot}$ in
steps of 0.2\,$M_{\odot}$ and the initial orbital period from $\log (P^{\rm
  i}/{\rm day})=1.5$ to 3.5 in steps of $0.1$. In the calculations, we
assume that a WD may explode as a SN Ia when its mass exceeds
1.378\,$M_{\odot}$ (\citealt{NTY84}). At the moment when  $M_{\rm WD}=1.378
M_{\odot}$, the companions are ascending either the first-giant branch
(FGB) or the asymptotic giant branch (AGB).  We then record the
status of the companions at this moment, i.e.  the secondary mass
$M_{\rm 2}^{\rm SN}$, the core mass $M_{\rm core}$ and the mass-loss
rate from the secondary $\dot{M}_{\rm 2}$. The definition of the core is the
same as that in \citet{HAN94} and \citet{MENG08}. We then estimate the
remaining mass-loss timescale of the secondary from
\begin{equation}
 t=\frac{M_{\rm env}}{|\dot{M}_{\rm 2}|}=\frac{M_{\rm 2}^{\rm
SN}-M_{\rm core}}{|\dot{M}_{\rm 2}|}.
\end{equation}
The mass-loss timescale here should be taken as the upper limit for
the spin-down timescale and we assume that it is the spin-down
timescale. The real mass-loss timescale could be different from
the value here since $\dot{M}_{\rm 2}$ is the value at the
moment when $M_{\rm WD}=1.378$ $M_{\odot}$, but the results here
should be reasonable, at least, to first order.




\section{RESULTS}\label{sect:3}

\begin{figure*}
\centerline{\includegraphics[angle=270,scale=.7]{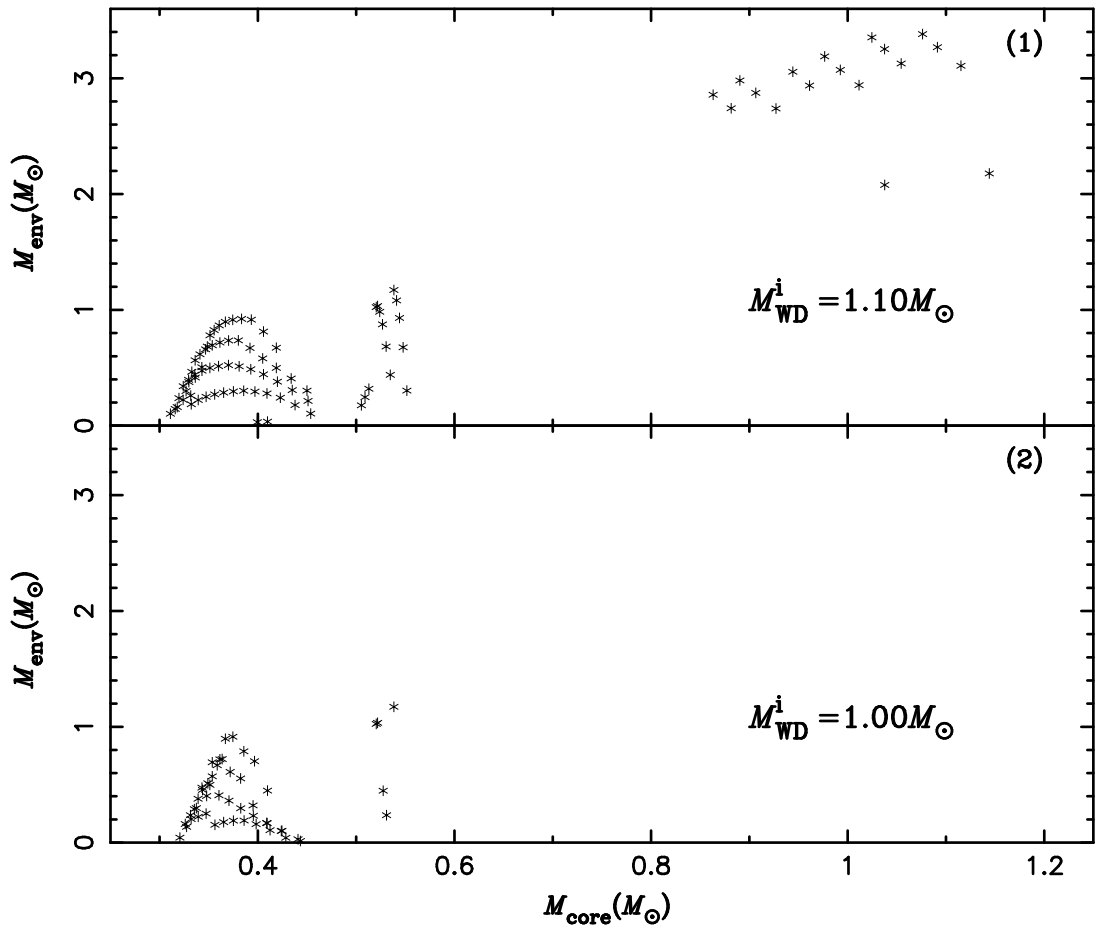}}
\caption{The core masses and the envelope masses of secondaries
for different initial WD mass when $M_{\rm WD}=1.378$
$M_{\odot}$.}\label{mcme}
\end{figure*}

Figure \ref{mcme} shows the core masses and the envelope masses of
secondaries at the moment when $M_{\rm WD}=1.378$ $M_{\odot}$. The
secondaries can be clearly divided into three groups: $M_{\rm
  core}<0.46 M_{\odot}$, $0.5 M_{\odot} < M_{\rm core} < 0.6
M_{\odot}$ and $M_{\rm core}>0.8 M_{\odot}$. For the first group, the
initial secondary mass $M_{\rm 2}^{\rm i}$ is smaller than $2.0
M_{\odot}$ and the core is composed of helium, while $M_{\rm 2}^{\rm
  i}$ in the second group ranges from $2.0 M_{\odot}$ to $2.4
M_{\odot}$ and the core is made of CO. $M_{\rm 2}^{\rm i}$ in the
third group is larger than $4.6 M_{\odot}$, up to $5.6 M_{\odot}$, and
the core is also made of CO (see also Figure 2 in
\citealt{CHENXF11}). The third group is very parameter dependent and
will disappear for lower values of $M_{\rm WD}^{\rm i}$ and $B_{\rm
  W}$ (\citealt{CHENXF11}), which indicates again that the results
here are conservative (i.e. include all possibilities). For those
with $M_{\rm 2}^{\rm i}\leq2.4 M_{\odot}$, the envelope mass at
the moment of $M_{\rm WD}=1.378$ $M_{\odot}$ is usually lower than
$1 M_{\odot}$, and may be as low as $0.01 M_{\odot}$, while for
those with $M_{\rm 2}^{\rm
  i}\geq4.6 M_{\odot}$, the envelope mass may be as high as $3.4
M_{\odot}$. If the secondary does not lose all of its envelope
before the supernova explosion, the envelope may be stripped off
by the supernova ejecta (\citealt{MAR00}; \citealt{MENGXC07}).
Therefore, after the interaction between the supernova ejecta and
the secondary, the final remnant would be a helium or CO WD for
the first and second groups, respectively. For the third group,
there could be thin helium layer above the CO core
(\citealt{MENG08}), and the final remnant could then be a CO WD or
a helium RG.

\begin{figure*}
\centerline{\includegraphics[angle=270,scale=.7]{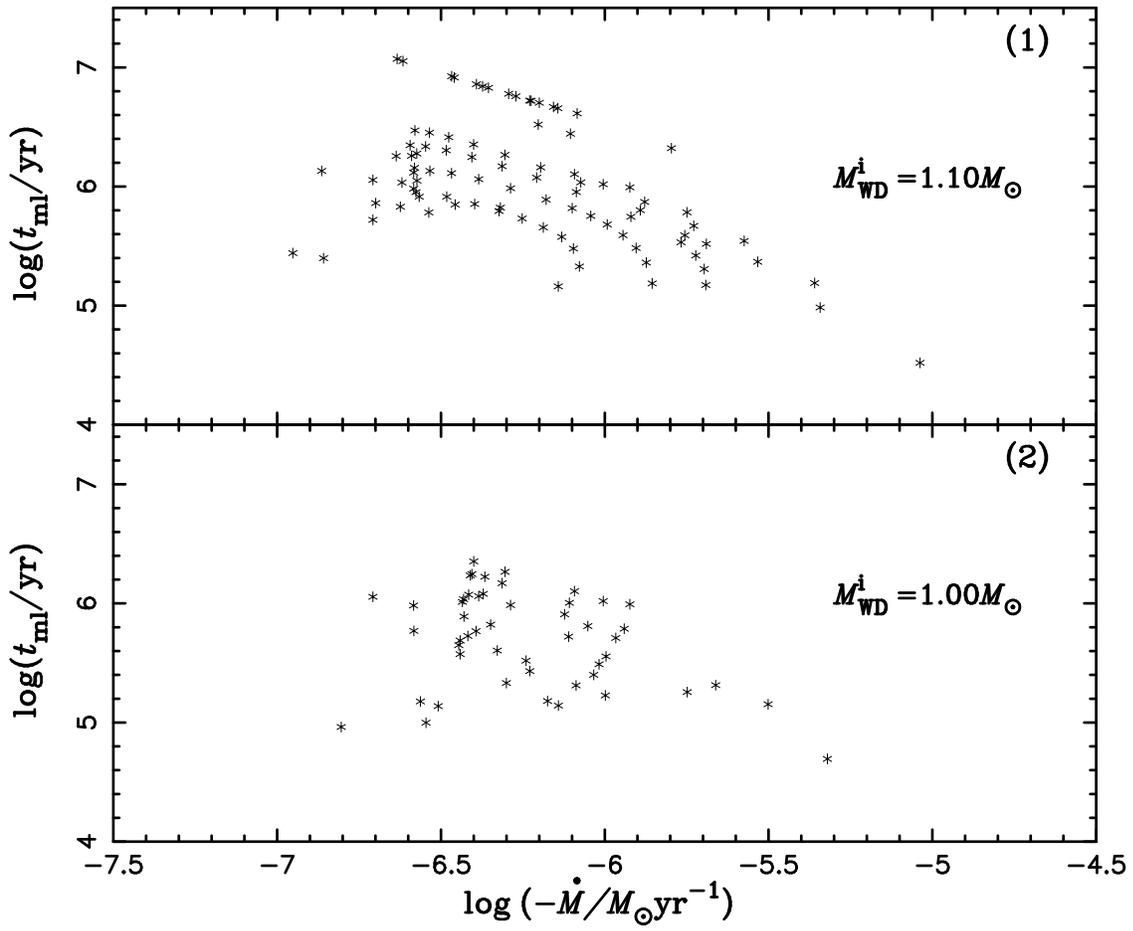}}
\caption{The mass-loss timecale as a function of mass-loss rate
of the secondary at the moment when $M_{\rm
WD}=1.378M_{\odot}$.}\label{mdottime}
\end{figure*}

Figure \ref{mdottime} shows the mass-loss timescale as a function
of mass-loss rate of the secondary at the moment when $M_{\rm
WD}=1.378M_{\odot}$. The mass-loss timescale here is still very
uncertain although the uncertainty is much smaller than that
quoted in \citet{DISTEFANO12}. Almost all systems have a mass-loss
timescale shorter than $10^{\rm 7} {\rm yr}$ and some are even
below $10^{\rm 5} {\rm yr}$. The mass-loss rate of the secondary
$|\dot{M}_{\rm 2}|$ is always larger than $10^{\rm -7} M_{\odot}
{\rm yr}^{\rm -1}$, even as high as $10^{\rm -5} M_{\odot} {\rm
yr}^{\rm -1}$, which indicates that the mass loss from some binary
systems at the moment when $M_{\rm WD}=1.378M_{\odot}$ is
dominated by the optically thick wind.

\begin{figure*}
\centerline{\includegraphics[angle=270,scale=.7]{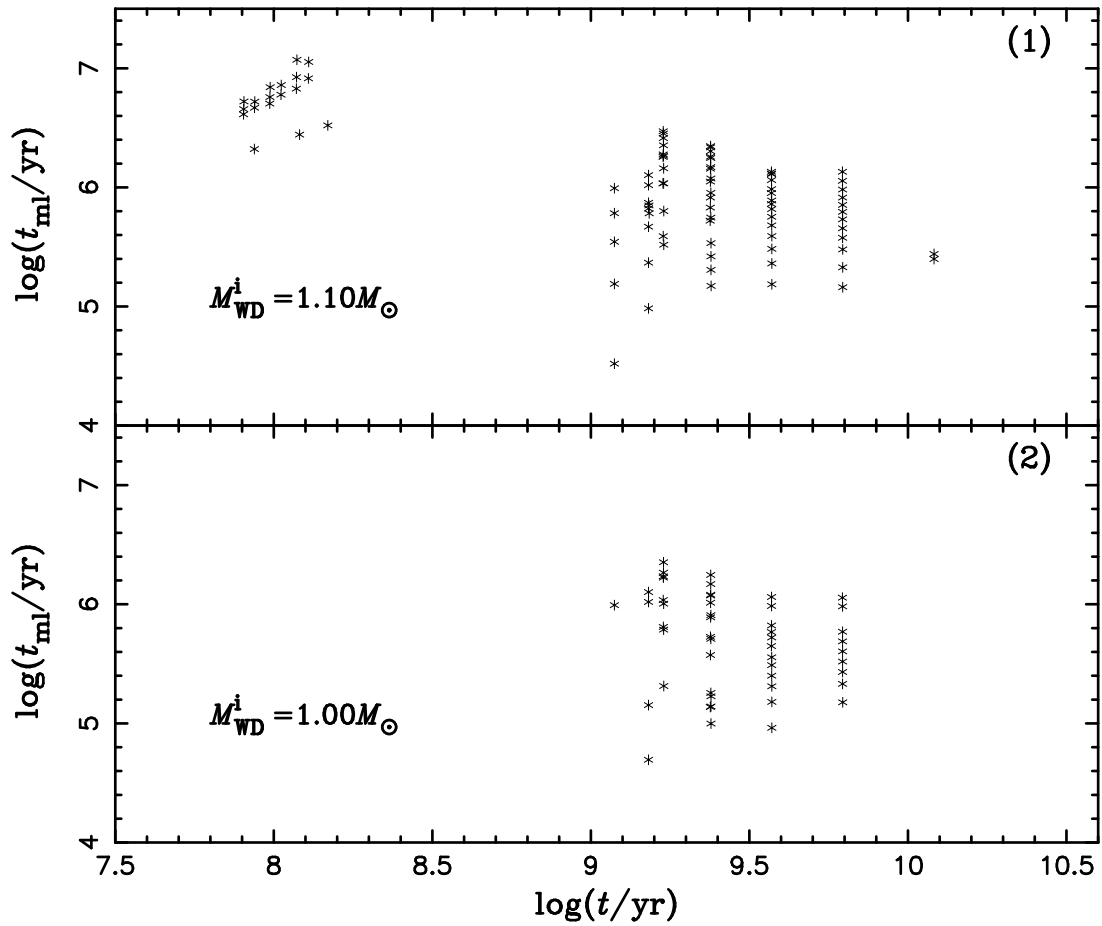}}
\caption{The mass-loss timescale as a function of the
evolutionary timescale of the secondary at the moment when $M_{\rm
WD}=1.378M_{\odot}$.}\label{sdtime}
\end{figure*}

Figure \ref{mdottime} shows that the mass-loss timescale is much
shorter than the evolutionary timescale of the secondary, (i.e.
from the ZAMS to the moment of $M_{\rm WD}=1.378M_{\odot}$). So,
we may take the evolutionary timescale of the secondary as the
delay time of the SNe Ia in the WD + RG channel. In Figure
\ref{sdtime}, we show the mass-loss timescale as a function of the
evolutionary timescale of the secondary. The evolutionary
timescale of the secondary covers a very large interval, i.e. from
less than $10^{\rm 8} {\rm yr}$ to more than $10^{\rm 10} {\rm
yr}$, which implies that the WD + RG channel may produce the
youngest, middle age and the oldest SNe Ia as well. In addition,
there is an interesting trend that the upper boundary of the
mass-loss timescale seemingly decreases with the evolutionary
timescale of the secondary; this is a consequence from the fact
that the star with a larger initial mass tends to leave a more
massive envelope when its WD companion has grown to
$1.378M_{\odot}$. This might indicate that it is more likely to
detect the signal of the CSM in a younger population.

 \section{DISCUSSION AND CONCLUSIONS}\label{sect:4}
 \subsection{Uncertainties}\label{sect:4.1}
\citet{YOON04b,YOON05} found that if the accretion rate onto a CO
WD is larger than $10^{\rm -7}$ $M_{\odot}$ ${\rm yr}^{\rm -1}$,
the WD may rotate strongly differentially (though this depends on
exactly how angular momentum is being redistributed within the
WD); this allows the WD to increase its mass up to $\sim 2$
$M_{\odot}$ before it can explode as a SN Ia or collapse to a
neutron star by electron capture. \citet{CHENWC09} and
\citet{HKSN12} puts these results into the canonical SD model to
try explain the origin of overluminous SNe Ia (SN 2003fg-like),
often referred to as `super-Chandra' SNe Ia. However, in the
simulations of \citet{YOON04b,YOON05}, the material that was
accreted was carbon/oxygen-rich material. Can we directly apply
these results to the case on accretion of hydrogen-rich material?
If so, are the results the same for hydrogen or helium burning on
a rotating WD and a non-rotating WD?  \citet{YOON04a} investigated
the effects of rotation on the helium-burning shell in an
accreting WD and found that helium burning in their rotating
models is much more stable than in the non-rotating model because
helium is ignited under much less degenerate conditions in the
rotating models. Then, how can this be applied to hydrogen
burning?

To explain the absence of hydrogen lines in the nebular spectrum
of SNe Ia (\citealt{LEO07}), \citet{JUSTHAM11} and
\citet{DISTEFANO11} proposed that the CO WD may experience a long
spin-down period to lose the angular momentum gained from the
accreted material .  This model may also explain the absence of
detectable surviving companions in SN Ia remnants
(\citealt{JUSTHAM11}; \citealt{DISTEFANO12}). However, there is
the more general question whether a WD can gain angular momentum
by accretion to spin itself up significantly. \citet{DISTEFANO12}
pointed to a sample of fast spinning WDs in close binary systems,
but mostly in so-called intermediate polars, which could mean that
a strong magnetic field may be necessary to spin up the WD.
However, it is still unclear whether a WD may effectively spin-up
by gas friction without magnetic field.



Assuming that the WD can be spun up by the accretion of hydrogen-rich
material and the hydrogen-rich material may be burned into helium and
then carbon/xoygen to increase the WD mass, when does the WD begin to
spin down? An accretion rate of $3\times10^{\rm -7}$ $M_{\odot} {\rm
  yr}^{\rm -1}$ was used as the threshold in \citet{CHENWC09}, while
it is $1\times10^{\rm -7}$ $M_{\odot} {\rm yr}^{\rm -1}$ in
\citet{HKSN12} and \citet{JUSTHAM11}. However, these values are based
on the accretion of carbon/oxygen-rich material from
\citet{YOON04b,YOON05}. It should be worth checking whether the
results in \citet{YOON04b,YOON05} can be generalized to the case of
the accretion of hydrogen-rich material, i.e. when does the WD begin
to spin down for the accretion of hydrogen-rich material? In this
paper, we just use $M_{\rm env}/|\dot{M}_{\rm 2}|$ to constrain the
spin-down timescale at the moment of $M_{\rm WD}=1.378 M_{\odot}$. If
the WD begins to spin down before this moment, the WD could be rigidly
rotating, and a very long spin-down timescale is expected
($\geq10^{\rm 9} {\rm yr}$, \citealt{HKSN12}). Then, the uncertainties
in our method would be small and could be neglected (0.1\%). On the
other hand, if the WD begins to spin down after $M_{\rm WD}=1.378
M_{\odot}$ has been reached, the time scale obtained here should be
taken as an upper limit. In addition, since $|\dot{M}_{\rm 2}|$
usually increases when the companion star climbs the FGB/AGB, the
$|\dot{M}_{\rm 2}|$ used here should generally be lower than the real
value. So, the timescale obtained in this paper should be taken as an
upper limit of the required spin-down timescale.

\begin{figure*}
\centerline{\includegraphics[angle=270,scale=.7]{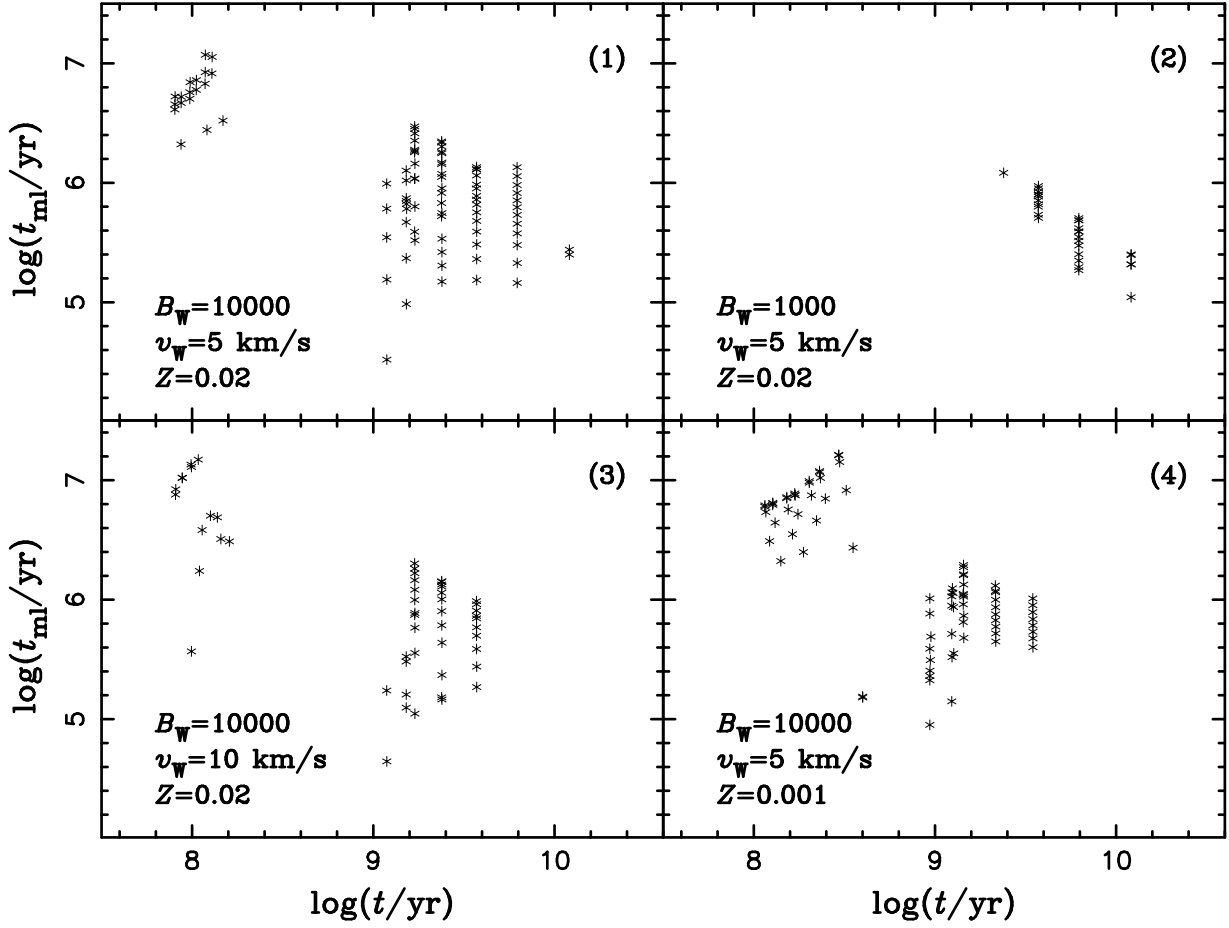}}
\caption{Similar to Figure \ref{sdtime} but for different values
of
  $B_{\rm W}$, $v_{\rm W}$ and different metallicities with
  $M_{WD}^{\rm i}=1.0 M_{\odot}$, where $Z=0.001$ is the lower
  metallicity limit allowed in the optically thick wind model
  (\citealt{HKSN12}).}\label{timemix}
\end{figure*}

So far, we only used one particular set for the poorly determined
paramters ($B_{\rm W}=10000$ and $v_{\rm W}=5$ ${\rm km}\,{\rm s}^{\rm
  -1}$) to model the problem. \citet{CHENXF11} have shown that the
parameter space leading to SNe Ia in the orbital period--secondary
mass plane decreases with decreasing $B_{\rm W}$ or increasing $v_{\rm
  W}$. In Figure \ref{timemix}, we show the effect of varying these
parameters, as well as metallicity, on the mass-loss timescale.
This figure shows that the upper boundary of the
  mass-loss timescale is not significantly affected by $v_{\rm W}$ and
  metallicity, but the influence of $B_{\rm W}$ is large. Generally,
a low $B_{\rm W}$ results in a lower wind mass-loss rate, which
means that the companion in a given binary system has a thicker
envelope when it fills its Roche lobe and it is more likely to
experience dynamically unstable mass transfer. So, the upper
boundary of the initial companion mass for SNe Ia moves to a lower
value with decreasing $B_{\rm W}$, in particular, the third group
in Fig. \ref{mcme} disappears (also see Figure 3 in
\citealt{CHENXF11}) and, as a consequence, the companion has a
less massive envelope when $M_{\rm WD}=1.378 M_{\odot}$, which
results in a lower mass-loss timescale. Actually, lower values of
$v_{\rm  W}$ and $Z$ may also lead to a slightly lower upper
boundary of the mass-loss timescale for a low initial WD mass, but
the effect is not as significant as for $B_{\rm W}$. For the lower
boundary of the mass-loss timescale, it is mainly determined by
the minimum envelope mass required to maintain effective accretion
onto the WD. In general, the mass of the minimum envelope mass is
similar for all values of $B_{\rm W}$, $v_{\rm W}$ and $Z$.

We also note that we did not consider the dispersion time of the
CSM ($10^{\rm 4}\sim10^{\rm 5}$ yr) and the simmering time
($10^{\rm 3}$) in the core of the WD before the supernova
explosion, since they can be neglected relative to the upper limit
of the mass-loss timescale.


For the discussion above, $10^{\rm 7}$ yr should be a conservative
upper limit of the spin-down timescale. On the theoretical side, the
spin-down timescale is very uncertain; it is mainly determined by the
timescale of the redistribution or loss of angular momentum in the WD,
whether it is rotating differentially or as a solid body. The
timescale for this angular momentum loss or redistribution is also
uncertain, but an upper limit of $10^{\rm 6}$ years has been claimed
from mapping the central density at ignition to the expected
nucleosynthesis (\citealt{YOON05}). This is consistent with our
constraints.

\subsection{The companions in SNR 0509-67.5}\label{sect:4.2}
In SNR 0509-67.5, \citet{SCHAEFER12} did not find a candidate for a
surviving companion, which they claimed favours the DD scenario.  On
the other hand, \citet{DISTEFANO12} argued that, if the spin-down
timescales is greater than $10^{\rm 5} {\rm yr}$, the donor star could
be too dim to be detectable at the time of explosion. In this paper,
we find that the spin-down timescale should usually be shorter than
$10^{\rm 7} {\rm yr}$, but also typically longer than $10^{\rm 5} {\rm
  yr}$. If the spin-down timescale is as long as $10^{\rm 7} {\rm yr}$
while the mass-loss timescale of the ex-companion is shorter then
this timescale, e.g. $10^{\rm 5} {\rm yr}$, the ex-companion in
SNR 0509-67.5 could have enough time to cool and to have a
luminosity lower than the observational limit (see Fig.~1 in
\citealt{DISTEFANO12}). However, one should keep in mind that the
spin-down timescale here has only been constrained for the WD + RG
channel.  It is unclear whether the spin-down timescale for the WD
+ MS channel is different from that for the WD + RG channel. In
addition, we do not know whether the region around the supernova
was empty or contained some CSM, in which case it is unclear
whether our calculations are directly relevant to the SNR. Hence,
it is too early to arrive at a conclusive conclusion concerning
the implications of the observations of \citet{SCHAEFER12}.
\section*{Acknowledgments}
We thank the anonymous referee for his/her constructive
suggestions. This work was partly supported by NSFC
(11003003,11273012), the Project of Science and Technology from
the Ministry of Education (211102) and Key Laboratory for the
Structure and Evolution of Celestial Objects, Chinese Academy of
Sciences.


\begin{thebibliography}{}
\bibitem[\protect\citeauthoryear{Alexander \& Ferguson}{1994}]{AF94}
Alexander, D. R., Ferguson J. W., 1994, ApJ, 437, 879
\bibitem[\protect\citeauthoryear{Boffin \& Jorissen}{1988}]{BOFFIN88}
Boffin, H. M. J. \& Jorissen, A., 1988, A\&A, 205, 155
\bibitem[\protect\citeauthoryear{Chen \& Tout}{2007}]{CHE07}
Chen, X., Tout, C.A., 2007, ChJAA, 7, 2, 245
\bibitem[\protect\citeauthoryear{Chen et al.}{2011}]{CHENXF11}
Chen, X., Han, Z., Tout, C.A., 2011, ApJ, 735, L31
\bibitem[\protect\citeauthoryear{Chen \& Li}{2009}]{CHENWC09}
Chen, W., Li, X., 2009, ApJ, 702, 686
\bibitem[\protect\citeauthoryear{Di Stefano et al.}{2011}]{DISTEFANO11}
Di Stefano, R., Voss, R., Claeys, J. S. W., 2011, ApJ, 738, L1
\bibitem[\protect\citeauthoryear{Di Stefano \& Kilic}{2012}]{DISTEFANO12}
Di Stefano, R., Kilic, M., 2012, ApJ, 759, 56
\bibitem[\protect\citeauthoryear{Dilday et al.}{2012}]{DILDAY12}
Dilday, B., Howell, D.A., Cenko, S.B. et al., 2012, Science, 337,
942
\bibitem[\protect\citeauthoryear{Eggleton}{1971}]{EGG71}
Eggleton, P.P., 1971, MNRAS, 151, 351
\bibitem[\protect\citeauthoryear{Gonz\'{a}lez-Hern\'{a}ndez et al.}{2009}]{HERNANDEZ09}
Gonz\'{a}lez-Hern\'{a}ndez J.I., Ruiz-lapuente P., Filippenko
A.V., Foley R.J., Gal-Yam A., Simon J.D., 2009, ApJ, 691, 1
\bibitem[\protect\citeauthoryear{Hachisu et al.}{1996}]{HAC96}
Hachisu, I., Kato, M., Nomoto, K., ApJ, 1996, 470, L97
\bibitem[\protect\citeauthoryear{Hachisu et al.}{1999a}]{HAC99a}
Hachisu, I., Kato, M., Nomoto, K., Umeda, H., 1999a, ApJ, 519, 314
\bibitem[\protect\citeauthoryear{Hachisu et al.}{2012}]{HKSN12}
Hachisu, I., Kato, M., Saio, H., Nomoto, K., 2012, ApJ, 744, 69
\bibitem[\protect\citeauthoryear{Han et al.}{1994}]{HAN94}
Han, Z., Podsiadlowski, P., Eggleton, P.P., 1994, MNRAS, 270, 121
\bibitem[\protect\citeauthoryear{Han et al.}{2000}]{HAN00}
Han, Z., Tout, C.A., Eggleton, P.P., 2000, MNRAS, 319, 215
\bibitem[\protect\citeauthoryear{Han \& Podsiadlowski}{2004}]{HAN04}
Han, Z., Podsiadlowski, Ph., 2004, MNRAS, 350, 1301
\bibitem[\protect\citeauthoryear{Hillebrandt \& Niemeyer}{2000}]{HN00}
Hillebrandt, W., Niemeyer, J.C., 2000, ARA\&A, 38, 191
\bibitem[\protect\citeauthoryear{Howell et al.}{2009}]{HOWEL09}
Howell, D.A. et al., 2009, arXiv: 0903.1086
\bibitem[\protect\citeauthoryear{Iben \& Tutukov}{1984}]{IT84}
Iben, I., Tutukov, A.V., 1984, ApJS, 54, 335
\bibitem[\protect\citeauthoryear{Iglesias \& Rogers}{1996}]{IR96}
Iglesias, C. A., Roger,s F. J., 1996, ApJ, 464, 943
\bibitem[\protect\citeauthoryear{Justham}{2011}]{JUSTHAM11}
Justham, S., 2011, ApJ, 730, L34
\bibitem[\protect\citeauthoryear{Kato \& Hachisu}{2004}]{KH04}
Kato, M., Hachisu I., 2004, ApJ, 613, L129
\bibitem[\protect\citeauthoryear{Kerzendorf et al.}{2009}]{KERZENDORF09}
Kerzendorf W.E., Schmidt B P, Asplund M, et al., 2009, ApJ, 701,
1665
\bibitem[\protect\citeauthoryear{Kerzendorf et al.}{2013}]{KERZENDORF13}
Kerzendorf W.E., Yong D., Schmidt B.P. et al., 2013, 774, 99
\bibitem[\protect\citeauthoryear{Leibundgut}{2000}]{LEI00}
Leibundgut, B., 2000, A\&ARv, 10, 179
\bibitem[\protect\citeauthoryear{Leonard}{2007}]{LEO07}
Leonard D.C., 2007, ApJ, 670, 1275
\bibitem[\protect\citeauthoryear{Maguire et al.}{2013}]{MAGUIRE13}
Maguire, K., Sullivan, M., Patat, F. et al., 2013, MNRAS.tmp, 2298
\bibitem[\protect\citeauthoryear{Marietta et al.}{2000}]{MAR00}
Marietta, E., Burrows, A., Fryxell, B., 2000, ApJS, 128, 615
\bibitem[\protect\citeauthoryear{Meng et al.}{2007}]{MENGXC07}
Meng, X., Chen, X., Han, Z., 2007, PASJ, 59, 835
\bibitem[\protect\citeauthoryear{Meng et al.}{2008}]{MENG08}
Meng, X., Chen, X., Han, Z., 2008, A\&A, 487, 625
\bibitem[\protect\citeauthoryear{Meng et al.}{2009}]{MENG09}
Meng, X., Chen, X., Han, Z., 2009, MNRAS, 395, 2103
\bibitem[\protect\citeauthoryear{Nomoto et al.}{1984}]{NTY84}
Nomoto, K., Thielemann, F-K., Yokoi, K., 1984, ApJ, 286, 644
\bibitem[\protect\citeauthoryear{Patat et al.}{2007}]{PAT07}
Patat, F. et al., 2007, Science, 317, 924
\bibitem[\protect\citeauthoryear{Perlmutter et al.}{1999}]{PER99}
Perlmutter, S. et al., 1999, ApJ, 517, 565
\bibitem[\protect\citeauthoryear{Reimers}{1975}]{REIMERS75}
Reimers, D., 1975, Mem. R. Soc. li\`{e}ge, 6i\`{e}me Serie, 8, 369
\bibitem[\protect\citeauthoryear{Riess et al.}{1998}]{RIE98}
Riess, A. et al., 1998, AJ, 116, 1009
\bibitem[\protect\citeauthoryear{Ruiz-Lapuente et al.}{2004}]{RUI04}
Ruiz-Lapuente, P. et al., 2004, Nature, 431, 1069
\bibitem[\protect\citeauthoryear{Schaefer \& Pagnotta}{2012}]{SCHAEFER12}
Schaefer, B. E., Pagnotta, A., 2012, Nature, 481, 164
\bibitem[\protect\citeauthoryear{Shappee et al.}{2013}]{SHAPPEE13}
Shappee, B.J.; Kochanek, C.S.; Stanek, K.Z., 2013, ApJ, 765, 150
\bibitem[\protect\citeauthoryear{Sternberg et al.}{2011}]{STERNBERG11}
Sternberg, A., Gal-Yam, A., Simon, J. D. et al., 2011, Science,
333, 856
\bibitem[\protect\citeauthoryear{Tout \& Eggleton}{1988}]{TOUT88}
Tout, C. A., \& Eggleton, P. P. 1988, ApJ, 334, 357
\bibitem[\protect\citeauthoryear{Tout \& Hall}{1991}]{TOUT91}
Tout, C. A. \& Hall, D. S., 1991, MNRAS, 253, 9
\bibitem[\protect\citeauthoryear{Wang \& Han}{2012}]{WANGB12}
Wang, B., \& Han, Z., 2012, NewAR, 56, 122
\bibitem[\protect\citeauthoryear{Webbink}{1984}]{WEB84}
Webbink, R.F., 1984, ApJ, 277, 355
\bibitem[\protect\citeauthoryear{Whelan \& Iben}{1973}]{WI73}
Whelan, J., Iben, I., 1973, ApJ, 186, 1007
\bibitem[\protect\citeauthoryear{Yoon et al.}{2004}]{YOON04a}
Yoon, S.-C., Langer, N., Scheithauer, S., 2004, A\&A, 425, 217
\bibitem[\protect\citeauthoryear{Yoon \& Langer}{2004}]{YOON04b}
Yoon, S.-C., Langer N., 2004, A\&A, 419, 623
\bibitem[\protect\citeauthoryear{Yoon \& Langer}{2005}]{YOON05}
Yoon, S.-C., Langer N., 2005, A\&A, 435, 967
\end{thebibliography}
\end{document}